\documentstyle[12pt,epsf]{article}
\epsfverbosetrue
\def\as{\alpha_{\rm S}}
\def\pT{p_{\rm T}}
\def\cO{{\cal O}}

\def\gev{{\rm\, GeV}}
\def\citenum#1{{\def\@cite##1##2{##1}\cite{#1}}}
\def\citea#1{\@cite{#1}{}}

\def\beq{\begin{equation}}
\def\eeq{\end{equation}}
\def\bea{\begin{eqnarray}}
\def\eea{\end{eqnarray}}

\def\bbbz{{\mathchoice {\hbox{$\sf\textstyle Z\kern-0.4em Z$}}
{\hbox{$\sf\textstyle Z\kern-0.4em Z$}}
{\hbox{$\sf\scriptstyle Z\kern-0.3em Z$}}
{\hbox{$\sf\scriptscriptstyle Z\kern-0.2em Z$}}}}
\def\npb#1#2#3{    {\it Nucl. Phys. }{\bf B#1} (19#2) #3}
\def\plb#1#2#3{    {\it Phys. Lett. }{\bf B#1} (19#2) #3}
\def\prd#1#2#3{    {\it Phys. Rev. }{\bf D#1} (19#2) #3}

\def\prl#1#2#3{    {\it Phys. Rev. Lett. }{\bf #1} (19#2) #3}

\def\rmp#1#2#3{    {\it Rev. Mod. Phys. }{\bf #1} (19#2) #3}
\def\zpc#1#2#3{    {\it Z. Phys. }{\bf C#1} (19#2) #3}
\def\mpla#1#2#3{   {\it Mod. Phys. Lett. }{\bf A#1} (19#2) #3}

\def\cpc#1#2#3{    {\it Comp. Phys. Commun. }{\bf #1} (19#2) #3}

\relax
\begin{document}
%
\begin{titlepage}
\noindent
DESY 92-135  \hfill  ISSN 0418-9833  \\
October 92\hfill                        \\[9ex]
\begin{center}
{\Large \bf Higher-Order QCD Corrections to          }   \\[1.4ex]
{\Large \bf Inclusive Particle Production
            in $p\bar p$ Collisions                  }   \\[11ex]

{\large F.M.\ Borzumati                             }    \\[1ex]
{\large B.A.\ Kniehl                                }    \\[1ex]
{\large G.\ Kramer                                  }    \\[1.5ex]
{\it II.\ Institut f\"ur Theoretische Physik
\footnote{Supported by the Bundesministerium f\"ur Forschung und
 Technologie, 05 5HH 91P(8), Bonn, FRG.   } }       \\
{\it Universit\"at Hamburg, 2000 Hamburg 50, Germany}      \\[21ex]

{\large \bf Abstract}
\end{center}
\begin{quotation}
Inclusive single-particle production cross sections have been
calculated
including higher-order QCD corrections.
Transverse-momentum and rapidity distributions are presented and
the scale dependence is studied.
The results are compared with experimental data from the CERN
$\rm Sp\bar pS$ Collider and the Fermilab Tevatron.
\end{quotation}
\end{titlepage}
\section{Introduction}

The inclusive production of single hadrons in hadronic collisions has
long been recognized as an important process for testing the
QCD-improved parton model (for reviews on the subject see, for instance,
ref.~\cite{COL}). The leading-order (LO) QCD formalism consists of
using the tree-level results for the hard cross sections, the one-loop
expression for the running coupling constant, and parton densities
(structure functions) and fragmentation functions generated by one-loop
evolution kernels. This theory provides a rather consistent description
of many large-momentum-transfer processes \cite{OWE}.
The tree-level parton-parton scattering cross sections that enter this
formalism have been calculated many years ago and an extensive
phenomenology has been developed and confronted with experimental data
\cite{COM,BAI}.

First results for next-to-leading order (NLO) corrections to
hard-scattering processes have been obtained
by Ellis et al.\ \cite{RKE} in 1980. However, these authors considered
only subprocesses with quarks in the initial state. Using quark
distribution functions defined in deep inelastic scattering
and fragmentation functions defined in one-particle inclusive
$e^+e^-$ annihilation, they found large corrections. Meanwhile, a
complete NLO calculation involving all parton-parton processes
contributing to hadron-hadron collisions has been presented by
Aversa at al.\ \cite{AVE}.

These results are requisite in order to perform a consistent
calculation of single-hadron production in the NLO formalism. This
involves NLO hard-scattering cross sections with two-loop $\as$,
two-loop-evolved structure functions and fragmentation functions. Only
under these circumstances are all four elements unambiguously defined.
Using NLO results for hard-scattering cross sections, structure and
fragmentation functions in combination, should reduce substantially
the scale dependence of the predictions for inclusive one-particle
cross sections, since the variation in one of these factors is
compensated by the others up to still missing higher-order terms.
To guarantee this compensation, a consistent choice of renormalization
and factorization schemes and scales is necessary in all four elements
entering the analysis. As mentioned above, NLO results for all
parton-parton subprocesses are available.

As is well known, structure and fragmentation functions cannot be
calculated yet from the QCD Lagrangian. For their construction one
has to resort to experimental data, in particular from total inclusive
deep-inelastic lepton-nucleon scattering in the case of the structure
functions and from one-particle inclusive $e^+e^-$ annihilation in the
case of the fragmentation functions. As for structure functions, such
analyses based on NLO evolution and all existing data have been carried
out by several groups, even including processes other than
deep-inelastic lepton-nucleon scattering \cite{MOR,MAR}; for a recent
review see \cite{JFO}.

Unfortunately, such analyses do not exist for the fragmentation
functions of quarks and gluons. Although the relevant NLO theory is
well known \cite{ALT} and inclusive single-particle distributions of
various hadrons have been measured in $e^+e^-$ annihilation as well as
in semi-inclusive deep-inelastic lepton-nucleon scattering, such a NLO
analysis has not been tackled yet. All the existing information on
fragmentation functions is based on LO evolution equations and also on
rather old data \cite{BAI,ANS}. Although this represents a serious
limitation to any NLO prediction of one-particle production in
hadron-hadron, photon-hadron, or lepton-hadron reactions, we consider
it appropriate to make the first step and to combine the available
results for the NLO parton-parton hard-scattering cross sections and
the NLO parton distribution functions to obtain more reliable
predictions as compared with the LO theory considered in the past.

Inclusive single-hadron production provides both advantages and
disadvantages as compared to jet production. A distinct hadron
constitutes a final state which is better defined than a jet event,
since the latter delicately depends on the details of the jet-finding
algorithm. Moreover, the definition of a jet becomes more problematic
at low energy. On the other hand, the fragmentation of the produced
parton renders the $p_T$ distributions quite steep and reduces
therefore the range experimentally accessible in the case of
single-hadron production. Accordingly, the experimental errors are
large in the high-$\pT$ range, where the cross sections become
considerably small, while the bulk of the data is accumulated at lower
values of $\pT$.

In this paper we shall present cross sections for
single-charged-hadron and single-$\pi^0$ production in $p\bar p$
reactions. Similar predictions for $\gamma p$ and low-$Q^2$ $ep$
processes will be considered in a separate communication. Experimental
data for charged one-particle and $\pi^0$ production at relatively low
centre-of-mass (CM) energies come from the CERN ISR $pp$
Collider \cite{ALP}. At higher energies, data from the UA1 \cite{UA1}
and UA2 \cite{BAN} Collaborations at the $\rm Sp\bar pS$ Collider
and from the CDF Collaboration \cite{PARA} at the Tevatron are
available. Since the QCD-improved parton model should work best for
the $p\bar p$ collider energy range, we shall compare our results
with data for the higher energies only. Specifically, we choose data at
$\sqrt{s}=540$ and $630\,$GeV from UA2 and CDF, respectively
and at $1.8\,$TeV from CDF.

The outline of this paper is as follows. In Sect.~2, we give a short
introduction of the formalism and fix our notation and input.
Results of our calculation are presented in Sects.~3 and 4, where we
also compare with data from $p\bar p$ colliders. Sect.~5 is reserved
for a discussion of the results and some concluding remarks.
\newpage
\section{Formalism, Notation, and Input}

In the LO QCD-improved parton model the single-particle cross section
can be expressed as a convolution of the LO parton-parton scattering
cross sections with the scale-dependent structure and fragmentation
functions. We fix the momenta of the hadron reaction by
\beq
 h_1(p_1)+h_2(p_2)\to h_3(p_3)+X\,,
\label{process}
\eeq
where $h_1$ and $h_2$ are the incoming hadrons and $h_3$ is the single
hadron in the final state. In LO the cross section is given by
\bea
\lefteqn{E_3\,{d^3\sigma^0\over d^3p_3} = } \nonumber  \\
& &
 \sum_{i,j,l} \int dx_1\int dx_2\int{dx_3\over x_3^2}
 \ F_i^{h_1}(x_1,M^2)\, F_j^{h_2}(x_2,M^2)
 \,D_l^{h_3}\left(x_3,M_f^2\right)
  \  p_3^0\, {d^3\sigma_{k_ik_j\to k_l}^0\over d^3p_3}
\label{crossect}
\eea
where the partonic momenta $k_i$, $k_j$ and $k_l$ are expressed
in terms of $p_1$, $p_2$ and $p_3$ as:
$$
k_i=x_1p_1,\quad k_j=x_2p_2, \quad k_l=p_3/x_3 \,.
$$
The indices $i,j,l$ run over gluons and $N_F$ flavours of quarks.
We assume $N_F=4$ throughout our calculation and neglect the influence
of the charm-quark threshold. $F_i^h(x,M^2)$ and $D_l^h (x,M_f^2 )$ are
the usual structure and fragmentation functions appropriate to partons
of type $i$ or $l$ inside hadron $h$, which depend on the scales $M$
and $M_f$ to be specified later. Finally, $d^3\sigma_{k_ik_j\to k_l}^0$
characterizes the process $i+ j\to l+X$ in
$\cO\left(\as^2(\mu^2)\right)$.

Thus, we have three scales altogether: $M$, the factorization scale of
the parton distributions $F_i^h$, $M_f$, the factorization scale of the
fragmentation functions $D_l^h$, and $\mu$, the renormalization scale
of the QCD coupling constant $\as$. In LO all three scales are left
undefined. Usually they are linked to the transverse momentum $\pT$ of
the produced hadron $h_3$. Since this choice of scale is rather
arbitrary, there is a large uncertainty of the theoretical predictions
at the LO level.

In NLO the inclusive cross section is written
as\footnote{In the following, we largely follow the notation of Aversa
 et al. \cite{AVE}.}
\bea
\lefteqn{E_3\,{d^3\sigma^0\over d^3p_3} =  } \nonumber  \\
 & &
   \sum_{i,j,l}\int dx_1\int dx_2\int{dx_3\over x_3^2}\,
  F_i^{h_1}(x_1,M^2)F_j^{h_2}(x_2,M^2)D_l^{h_3}\left(x_3,M_f^2\right)
                                           \nonumber  \\
&  &
  \quad    \times
  \left[{1\over v}
  \left({d\sigma^0 _{k_ik_j\rightarrow k_l} (s,v)\over dv}\right)
  \delta(1-w)+{\as(\mu^2)\over2\pi}\,
  K_{k_ik_j\to k_l}\left(s,v,w;\mu^2,M^2,M_f^2\right)\right],
\label{mastereq}
\eea
where $v$ and $w$ can be expressed in term of the partonic variables
$s$, $t$ and $u$ as:
$$
 v=1+t/s\,,\qquad w=-u/(s+t)\,.
$$
In turn, $s$, $t$ and $u$ are related to the hadronic variables
$S =(p_1+p_2)^2$, $T=(p_1-p_3)^2$ and $U=(p_2-p_3)^2$ by:
$$
 s=x_1x_2S\,,\qquad
 t={x_1\over x_3}\,T\,,\qquad
 u={x_2\over x_3}\,U\,,
$$
The factors
$K_{k_ik_j\to k_l}\left(s,v,w;\mu^2,M^2,M_f^2\right)$
are the NLO corrections.

The calculation of the correction terms $K_{k_ik_j\to k_l}$ is fully
described in \cite{AVE}. It is based on the results of Ellis and Sexton
\cite{SEX} for the matrix elements squared for all $(2\to2)$ and
$(2\to3)$ parton scattering subprocesses calculated to
$\cO\left(\as^3\right)$. The $K_{k_ik_j\to k_l}$ functions are plagued
by infrared and collinear divergences, which appear as poles in the
parameter $\varepsilon=(4-n)/2$, with $n$ being the dimensionality of
space-time. The cancellation of the infrared singularities, which are
$\propto1/\varepsilon^2$, is explained in \cite{AVE}. The left-over
$1/\varepsilon$ singularities associated with ingoing parton lines are
absorbed into the bare distribution functions $F_i^h(x)$, which renders
them dependent on the scale $M$. Similarly, the $1/\varepsilon$
singularities from outgoing parton lines are absorbed into the bare
fragmentation functions $D_l^h(x)$, so that they become dependent on
the scale $M_f$.

After all these cancellations and absorptions have been done, the
single-hadron inclusive cross section has the form (3) with the
correction terms
$K_{k_ik_j\to k_l}$ being given by \cite{AVE}
\bea
\lefteqn{
  K_{k_ik_j\to k_l}\left(s,v,w;\mu^2,M^2,M_f^2\right)=
        \hspace*{3.0cm}             }      \nonumber \\[1.2ex]
& &
{1\over v}\left({d\sigma^0_{k_ik_j\to k_l}(s,v)\over dv}\right)
 \ \ \left\{\left(c_1+\tilde c_1\ln{s\over M^2}
  +\tilde{\tilde c}_1\ln{s\over M_f^2}+
   \hat c_1\ln{s\over\mu^2}\right) \delta(1-w)
                                  \right.   \nonumber  \\
& &
  \hspace*{3.0cm} \left.
 +\left(c_2+\tilde c_2\ln{s\over M^2}
 +\tilde{\tilde c}_2\ln{s\over M_f^2}\right)
 \left({1\over1-w}\right)_+
 +c_3\left[{\ln(1-w)\over1-w}\right]_+\right\}
                                            \nonumber \\[1.2ex]
& &
   \hspace{8cm} +\ K^\prime_{k_ik_j\to k_l}(s,v,w)\,,
\label{kfactor}
\eea
where the $K^\prime_{k_ik_j\to k_l}(s,v,w)$ functions are regular in
the limit $w\to1$. The coefficients $c_i$, $\tilde c_i$,
$\tilde{\tilde c}_i$, and $\hat c_i$ may be found in \cite{AVE} for the
various subprocesses. The $K^\prime_{k_ik_j\to k_l}$ functions
are given by rather lengthy expressions and have been communicated to
us by the authors of \cite{AVE}.

The form of the various coefficients and the $K^\prime_{k_ik_j\to k_l}$
functions in (\ref{kfactor}) depend on the choice of the finite
corrections $f_{ij}(x)$ and $d_{ij}(x)$ ($i,j=q,g$) to structure and
fragmentation functions. These finite corrections have to be adjusted
according to the selected factorization scheme. The authors of
\cite{AVE} advocate a scheme where all $f_{ij}$ and $d_{ij}$ are kept
different from zero in the form specified in \cite{AVE}. However, this
alternative reaches beyond the scope of present structure-function
analyses, since it affects higher orders relative to the NLO level, and
we shall not follow it. The popular DIS and $\overline{\rm MS}$ schemes
require $f_{ij}=d_{ij}=0$, except for $f_{qq}$ and $d_{qq}$, which are
finite and different for the two schemes. The expressions for
$f_{qq}$ and $d_{qq}$ listed in \cite{AVE} correspond to structure and
fragmentation functions in the $\overline{\rm MS}$ scheme. Since
fragmentation functions have been determined only to LO, the
$\overline{\rm MS}$ scheme is appropriate here ($d_{qq}\ne0$).
As for structure functions, results are available in both
$\overline{\rm MS}$ and DIS schemes. Depending on the choice of scheme,
the appropriate forms of $f_{qq}$ must be selected in the NLO
corrections to the parton-level cross sections.

As already mentioned in the introduction, we shall employ the
fragmentation functions by Baier et al. \cite{BAI} and Anselmino
et al. \cite{ANS}, which appears to be the best choice for the time
being. Reference \cite{ANS} provides us with functions for the
fragmentation of $u$ and $d$ quarks into $\pi^\pm$ and $K^\pm$ as a
function of the scale $M_f$. They are obtained from fits to EMC
deep-inelastic lepton scattering data. We adopt the fragmentation
functions of the gluon into $\pi^{\pm,0}$ from \cite{BAI},
with $D_g^{\pi^+}=D_g^{\pi^-}=D_g^{\pi^0}$. These, as well as the
$u$ and $d$ quark fragmentation functions have a reference scale
$Q_0 =5\,$GeV. Following \cite{ORG}, we identify the fragmentation
function of the $s$ quark into a given pion with the one of the
first-family quark that is not a ground-state constituent of the
respective pion, e.g., $D_s^{\pi^+}=D_d^{\pi^+}$, etc.
We proceed similarly in the case of $c$ quarks. The fragmentation
functions into kaons emerge from those into pions through
multiplication by heuristic reduction factors \cite{ANS}.

The proton structure function parametrizations used in our
calculcations are taken from the package by K. Charchu\l a \cite{CHA}.
For the central results and, in particular, for the comparison with
experimental data we use the Morfin-Tung set MT-B1 \cite{MOR},
either in the DIS or in the $\overline{\rm MS}$ version.

In the next sections, we shall present our results in two steps.
In the first step, we assume
$D_i^h (x_3,M_f^2)= \delta(1-x_3)$ for all hadrons $h$ and partons $i$.
In this way, we study the cross sections for the production of single
partons. Although these cross sections cannot be compared with
experimental data---a fragmentation function $\sim\delta(1-x_3)$ is a
reasonable approximation only for heavy quarks---, they are simpler to
calculate and reveal several features and dependences that are also
relevant for the complete single-hadron cross sections considered in
the second step. We emphasize that these cross sections must not be
compared with experimental jet production cross sections, since they
lack a recombination algorithm indispensable for defining jets.
\newpage
\section{Results: Single-Parton Inclusive Cross Section}

%
\begin{figure}[p]
\epsfxsize=16cm
\epsfbox{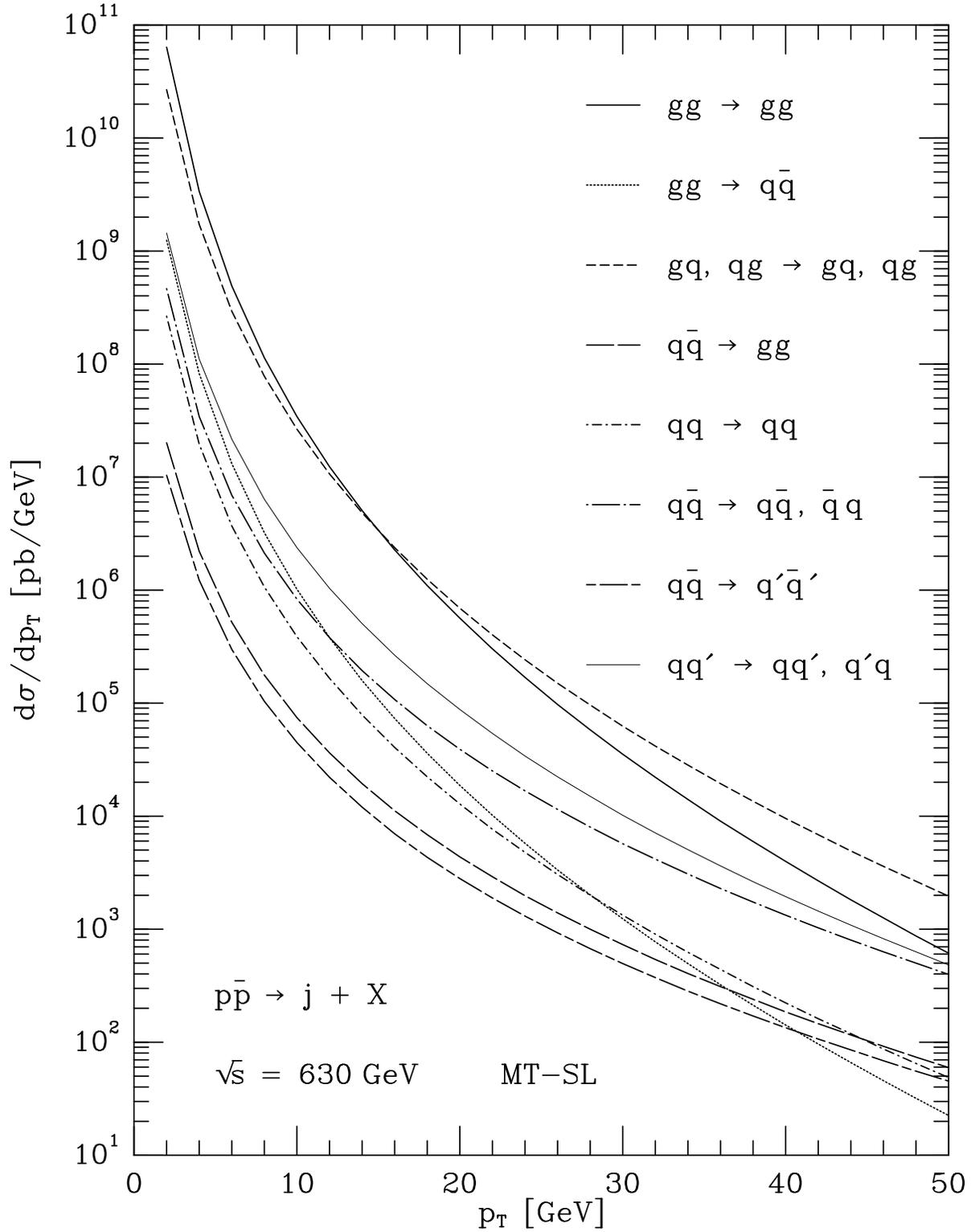}
\caption[f1]{\small{$\pT$ distributions of the various LO channels at
 $\sqrt{s}=630\,$GeV evaluated with the LO set SL of the
 Morfin-Tung structure functions.     }}
\label{pchannel}
\end{figure}
In this section we approximate all fragmentation functions by $\delta$
functions. This leads to single-parton inclusive cross sections. For a
first orientation, we disentangle the various parton-parton channels
that contribute in LO to the hard-scattering cross section, viz.\
\begin{itemize}
\item[1)] $gg\to gg$,
\item[2)] $gg\to q\bar q$,
\item[3)] $gq\to gq, qg\ \ $ and $\ \ qg\to qg , gq$,
\item[4)] $q\bar q\to gg$,
\item[5)] $qq\to qq$,
\item[6)] $q\bar q\to q\bar q , \bar qq$,
\item[7)] $ q\bar q\to q^\prime\bar q^\prime$,
\item[8)] $qq^\prime\to qq^\prime , q^\prime q$,
\end{itemize}
where $q$ and $q^\prime$ denote different quark flavours.
In Fig.~\ref{pchannel}, this is done for $d\sigma/d\pT$ at
$\sqrt s=630\gev$ using the LO set SL of the Morfin-Tung structure
functions \cite{MOR}. Here and in the following figure we choose
$\mu=M=\pT$. We observe that at small $\pT$ the two channels
1) and 3)
dominate by more than one order of magnitude, whereas at larger $\pT$
($\pT>30\gev$) also the pure quark channels 6) and 8) contribute
significantly. Similar observations have been made in the
past \cite{HUM}. It is clear that the magnitude of the cross
sections with gluonic initial states are highly sensitive to the gluon
content of the proton (antiproton), whose small-$x$ behaviour is
still not well known. Of course, the relative size of the various
channels in Fig.~\ref{pchannel} is subject to change when realistic
fragmentation functions are taken into account.

\begin{figure}[tb]
\vspace{12.4cm}
\caption[f2]{\small{Influence of typical NLO structure functions
 on the LO calculation of $d^3\sigma/dy\,d^2p_T$ at $\sqrt{s}=630\,$GeV
 and $y=0$. The curves are normalized to the LO calculation with the
 MT-SL set.               }}
\label{sfdepen}
\end{figure}
Next, we study how our results are influenced when different sets
of structure functions are used. Toward this end, we calculate
$d^3\sigma/dy\,d^2\pT$ in the Born approximation at $\sqrt s=630\gev$
and $y=0$ as a function of $\pT$ using in turn the structure-function
sets MT-B1 (DIS), MT-B1 ($\overline{\rm MS}$) \cite{MOR} and
MRSn-D0 ($\overline{\rm MS}$), MRSn-D$-$ ($\overline{\rm MS}$)
\cite{MRSN}, which are all NLO structure functions. For the QCD
coupling $\as$ we employ the usual one-loop formula.
In Fig.~\ref{sfdepen}, we show these cross sections normalized with
respect to the value obtained when the MT-SL set is used. Strictly
speaking, only the latter should be combined with the LO
hard-scattering cross sections because it is a genuine LO fit; all the
others correspond to NLO analyses of present data.

We observe that the variation caused by replacing the LO set by the
various NLO sets does not exceed some 30\% for $\pT>5\gev$. The DIS
and $\overline{\rm MS}$ versions of MT-B1 differ by only 10\%
relative to each other. We may consider this as an upper bound on the
scheme dependence, since the partial compensation that is expected to
take place at NLO due to scheme-dependent terms is not in effect here.
The MRSn-D0 and MRSn-D$-$ fit also recent NMC and CCFR \cite{NMCCCFR}
data which extend to smaller values of $x$ than previously
tested. The results differ significantly at low $\pT$. In particular,
the D0 and D$-$ sets differ drastically from each other. This may be
understood by recalling that the gluon-initiated subprocesses dominate
at low $\pT$; see Fig.~\ref{pchannel}. The variation is considerably
reduced at larger $\pT$ ($\pT>30\gev$), where also the channels (6) and
(8) contribute significantly. These features are expected to survive at
NLO.

\begin{figure}[tb]
\vspace{12.4cm}
\caption[f3]{\small{$d^3\sigma/dy\,d^2p_T$ in NLO normalized
  to the corresponding calculation with the NLO terms in the
  hard-scattering cross sections omitted, for $\sqrt{s}=630\,$GeV,
  $y=0$, and scales set to $\pT/2$, $\pT$, and $2 \pT$.
  DIS and $\overline{\rm MS}$ results are compared.       }}
\label{ratio}
\end{figure}
To investigate the influence of the higher-order terms in the
hard-scattering cross sections, we calculate $d^3\sigma/dy\,d^2\pT$
at $\sqrt s=630\gev$ and $y=0$ as a function of $2\pT/\sqrt s$ with
and without these terms and plot the ratio in Fig.~\ref{ratio}. In both
numerator and denominator we use the same structure functions,
namely the DIS or $\overline{\rm MS}$ versions of MT-B1, and the
two-loop expression for $\as$. The renormalization scale
$\Lambda$ of $\as$ is adjusted according to the structure
functions. We do this for the three choices $\mu=M=M_f=\pT/2$, $\pT$,
and $2\pT$. From Fig.~\ref{ratio} we see that this ratio is
independent of $\pT$, except for small $\pT$. Its value ranges between
1.3 and 2.1, which demonstrates the importance of the NLO corrections to
the parton-level cross sections. The scheme dependence DIS versus
$\overline{\rm MS}$ is negligible. Note that, in contrast to Fig.~2,
the same structure functions are used in both numerator and
denominator, so that variations due to changes of scheme or set largely
cancel out. On the other hand, we observe a strong scale dependence,
which is mainly due to the denominator as will become apparent below.

\begin{figure}[p]
\vspace{20cm}
\caption[f4]{\small{Scale dependence of $d^3\sigma/dy\,d^2p_T$ for
  $\sqrt{s}=630\,$GeV, $y=0$, and a) $\pT=31.5\,$GeV, b)
  $\pT=94.5\,$GeV. The solid and dot-dashed lines represent the NLO
  results with variable and fixed $M_f$, respectively; the dashed lines
  represent the LO result with NLO structure functions and $\as$.
                                        }}
\label{scalpart}
\end{figure}
The scale dependence of numerator and denominator of the ratio
considered in Fig.~3 is explicitly studied in Fig.~\ref{scalpart}a
and \ref{scalpart}b for $\pT$ fixed at $31.5\,$GeV and $94.5\,$GeV,
respectively. We set the three scales $\mu,M$ and $M_f$ equal to $\xi
\pT$ and we let $\xi$ vary between $0.125$ and $8$. The higher values
of $\xi$ in Fig.~\ref{scalpart}b allow for scales exceeding the CM
energy. For the results obtained in these figures, the DIS scheme
is adopted in connection with the appropriate MT-B1 set. Obviously, the
LO cross section (dashed line) exhibits a stronger dependence on
the scale factor $\xi$ as compared to the NLO result (solid line). The
more rapid decrease of the LO cross section for increasing $\xi$
explains the mild increase of the ratio plotted in Fig.~\ref{ratio}.
The NLO cross section is particularly insensitive to $\xi$ for $\xi>1$.
This holds true independently of $\pT$, although the absolute cross
section strongly varies with $\pT$; the cross sections shown in
Figs.~\ref{scalpart}a and b differ by four orders of magnitude. The
scale variation of the NLO cross section is well bounded in contrast to
the LO cross section, which decreases monotonically with increasing
$\xi$. However, one should keep in mind that the $M_f$ dependence of
the NLO result is not compensated as long as $\delta$-function-type
fragmentation is assumed. In this somewhat pathological situation it
would be more appropriate not to vary $M_f$. For comparison, we have
included the respective results in Fig.~\ref{scalpart} also (dot-dashed
curves).
\newpage
\section{Results: Single-Hadron Inclusive Cross Section.}

In this section we shall present our results for the inclusive
production of charged hadrons and of single $\pi^0$'s. For the
calculation of these cross sections we have employed the
fragmentation functions already mentioned in Sect.~2. For the
comparison with experimental data we have chosen the structure
function MT-B1 in the DIS scheme. Collider data for charged
hadrons ($h=(h^+ +h^-)/2$) come from the UA2 and CDF
Collaborations \cite{BAN,PARA}.

\begin{figure}[p]
\epsfxsize=14.8cm
\epsfbox{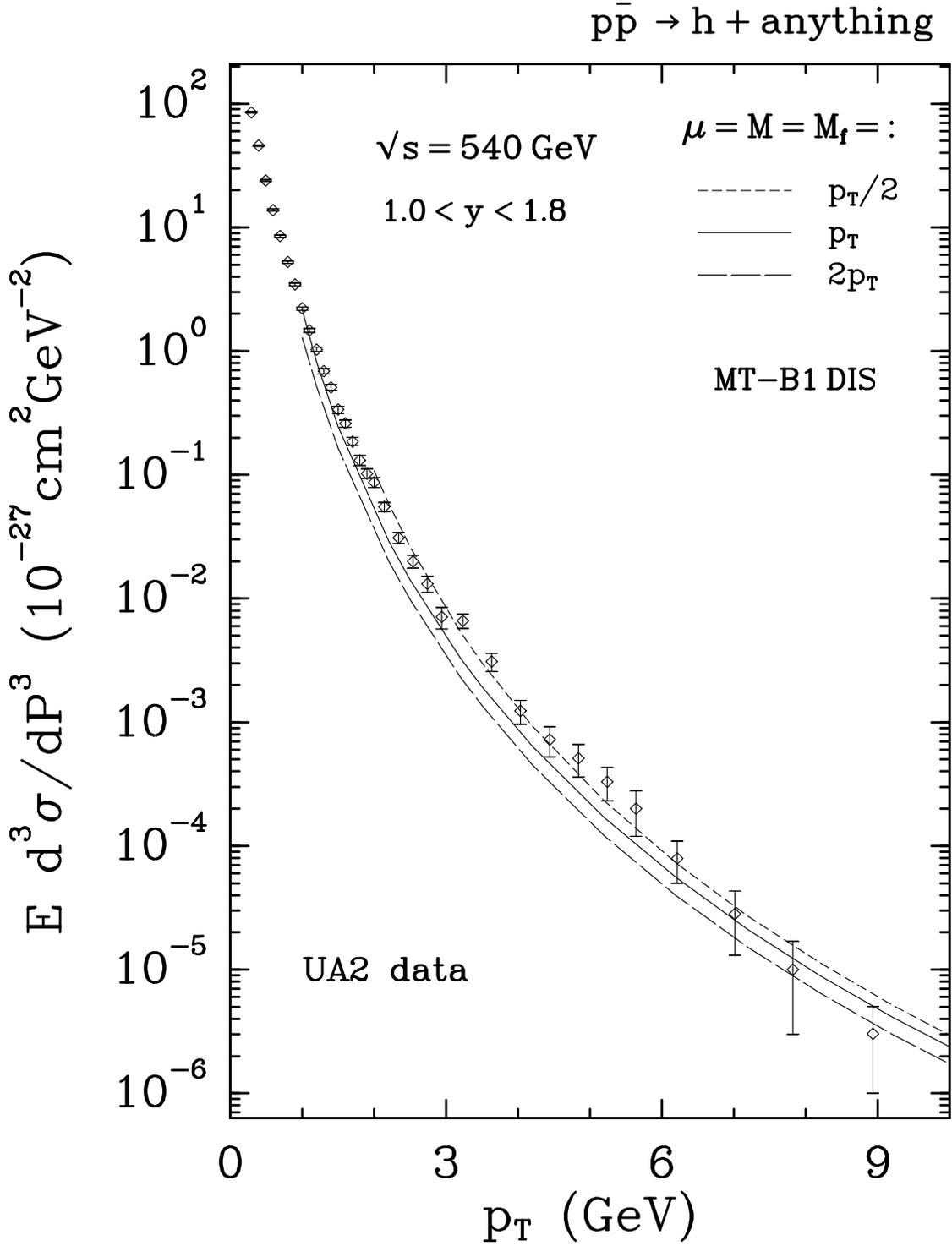}
\caption[f5]{\small{Inclusive charged-hadron production
  cross section for $\sqrt{s}=540\,$GeV, averaged over the
  rapidity range $1.0<y<1.8$. The short-dashed, solid and long-dashed
  lines correspond to the full NLO predictions for scales $\mu$, $M$
  and $M_f$ simultaneously set equal to $\pT/2$, $\pT$ and $2\pT$.
  For comparison, the UA2 data taken at the same CM energy and in the
  same rapidity range are also shown.                        }}
\label{UA2cha}
\end{figure}
In Fig.~\ref{UA2cha} we compare the charged-hadron inclusive cross
section as measured by UA2 at $\sqrt{s}=540\,$GeV, for $1.0<y<1.8$,
with the results we obtain when the three different scales $\mu$, $M$
and $M_f$ are simultaneously set equal to $\pT/2$, $\pT$ and $2 \pT$.
This scale dependence is rather moderate and we shall discuss later on
how it compares with the scale dependence in the LO cross section. Up
to this uncertainty, this is an absolute prediction of the cross
section. We observe that the prediction with scales equal to $\pT$ is
in adequate overall agreement with the data for $\pT$ up to
$10\,$GeV. The agreement is less satisfactory at small $\pT$, where
higher order corrections may be appreciable. In addition, somewhere in
this region non-perturbative effects come in.

\begin{figure}[p]
\epsfxsize=14.8cm
\epsfbox{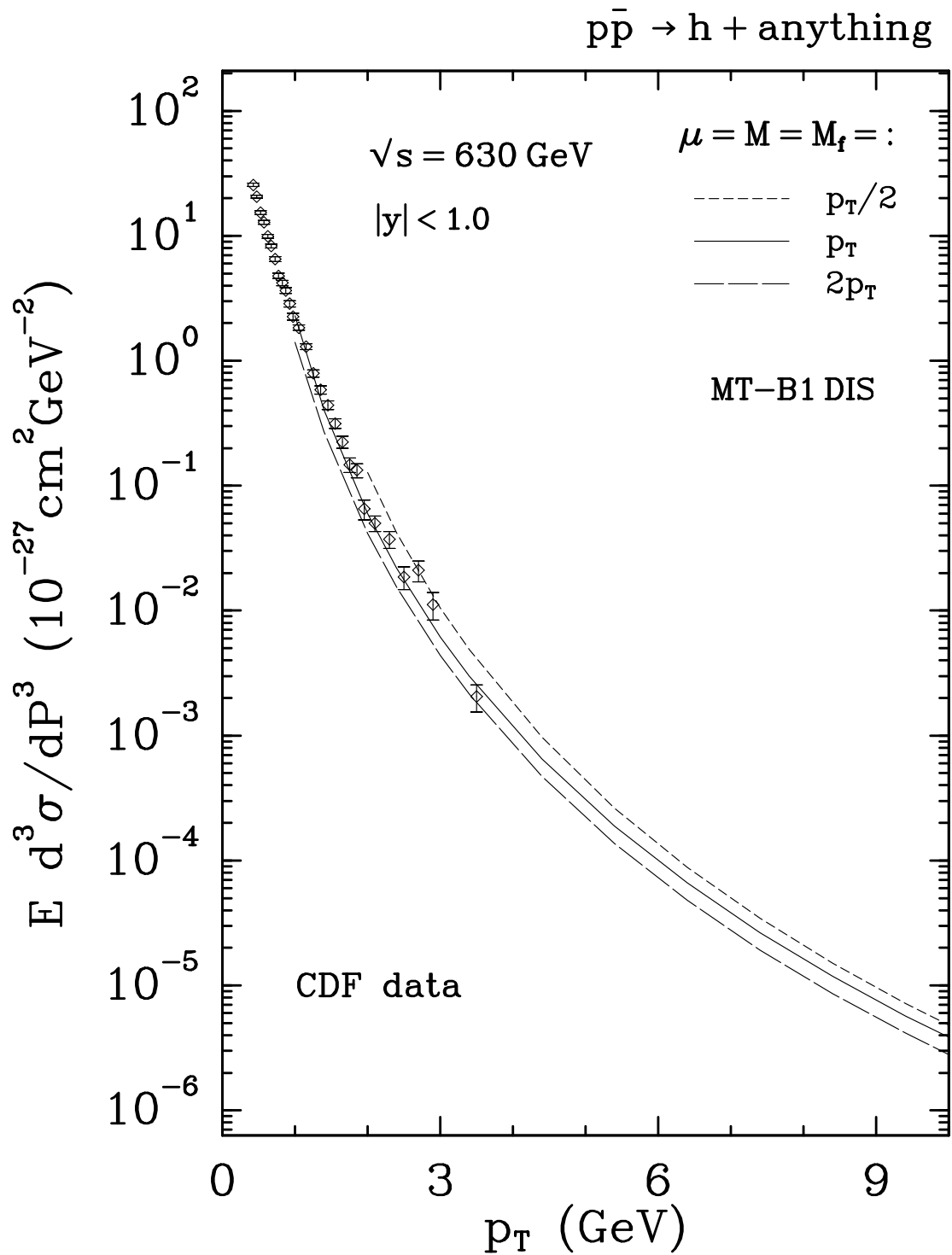}
\caption[f6]{\small{Same as in Fig.~\ref{UA2cha} for
  $\sqrt{s}=630\,$GeV and rapidity range $\vert y \vert < 1.0 $.
  The theoretical results are here compared with data obtained by
  the CDF Collaboration.    }}
\label{CDFonea}
\end{figure}
\begin{figure}[p]
\epsfxsize=14.8cm
\epsfbox{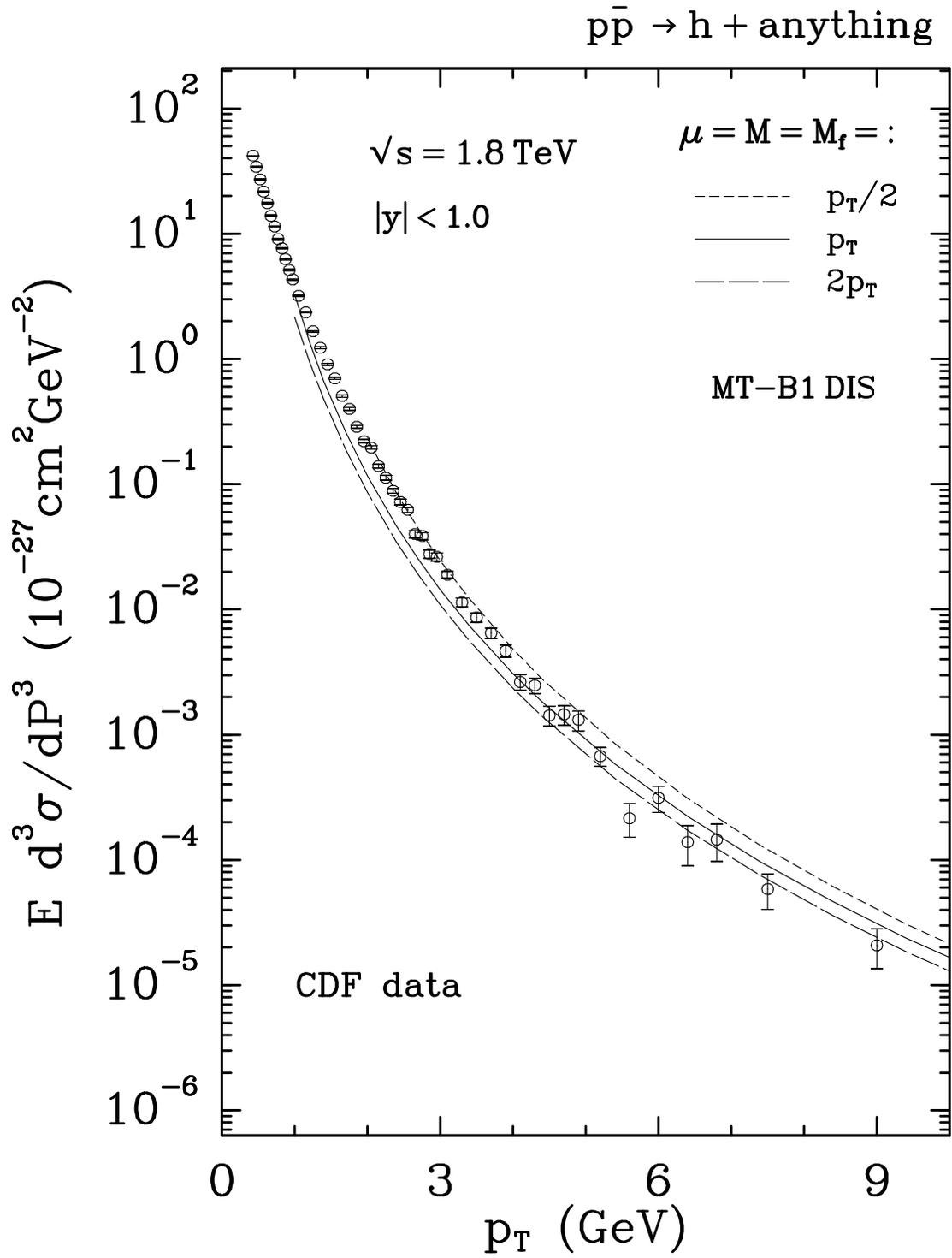}
\caption[f7]{\small{Same as in Fig.~\ref{CDFonea} for
  $\sqrt{s}=1.8\,$TeV.      }}
\label{CDFtwoa}
\end{figure}
Similar conclusions can be drawn for the comparisons shown in
Figs.~\ref{CDFonea} and \ref{CDFtwoa} with CDF data at
$\sqrt{s}=630\,$GeV and $\sqrt{s}=1.8\,$TeV, respectively. In both
cases, the rapidity is averaged over the interval
$-1 < y < 1$. The $1.8\,$TeV data at large $\pT$ are again predicted
best with scales equal to $\pT$. The same seems to be true for the
$630\,$GeV data, although the $\pT$ range is in this case too small to
allow a firm statement.

Looking at the energy dependence, we note that $\sqrt{s}$ increases
by a factor 3 as we go from Figs.~\ref{UA2cha},\ref{CDFonea} to
Fig.~\ref{CDFtwoa}. We observe that at small $\pT$, below $3\,$GeV, the
cross section is independent of $\sqrt{s}$, whereas in the large-$\pT$
range the cross section increases with increasing energy, i.e. it shows
the characteristic flattening of the $\pT$ distribution originating
from the hard-scattering cross section. This behaviour of the data is
borne out by the theory quite well although at small $\pT$ we would not
expect such a good agreement.

\begin{figure}[p]
\epsfxsize=14.8cm
\epsfbox{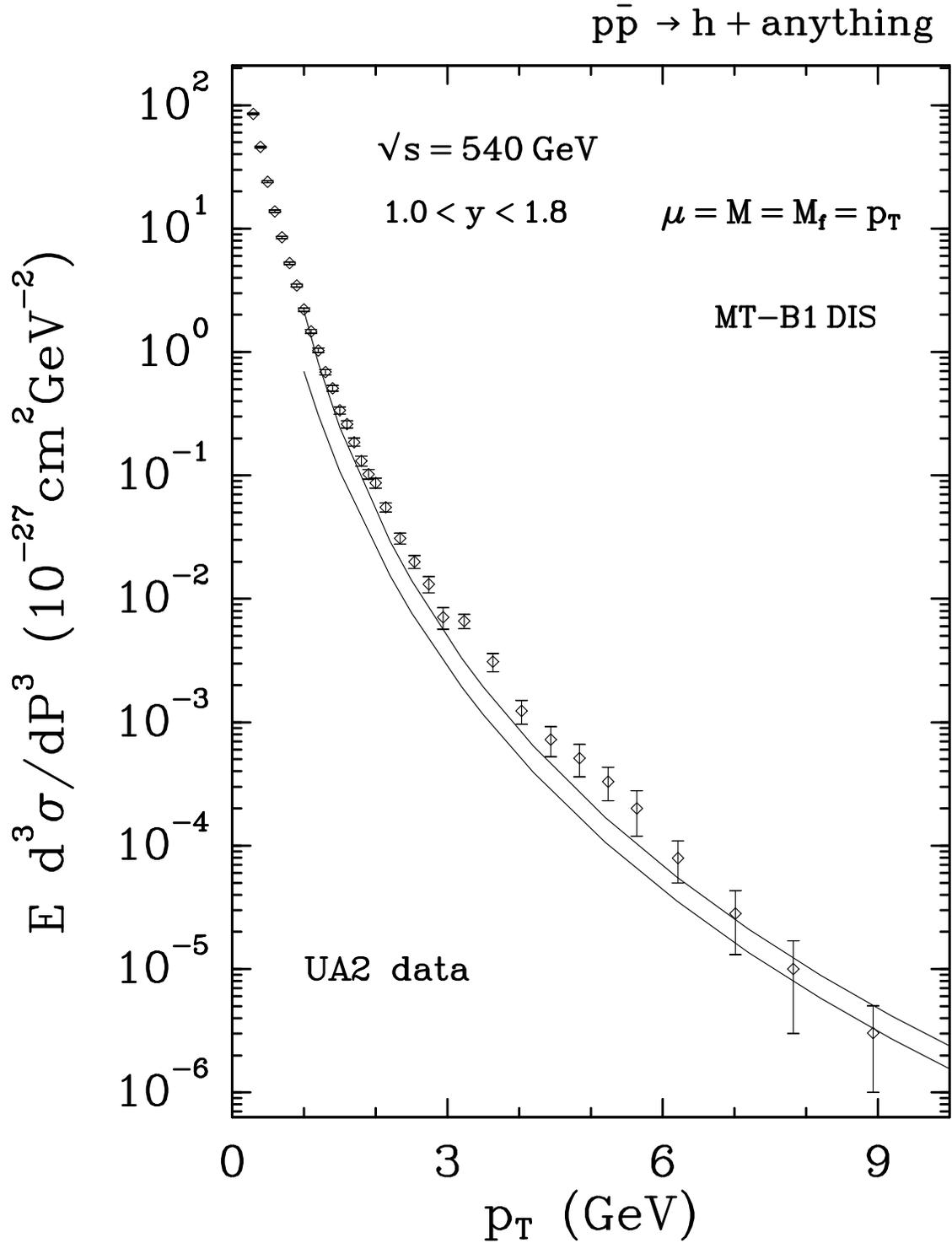}
\caption[f8]{\small{The experimental data shown in
  Fig.~\ref{UA2cha} are here compared to the NLO theoretical
  result (upper curve) and the LO one (lower curve) as obtained for
  structure functions and $\as$ kept at the NLO level.      }}
\label{UA2chb}
\end{figure}
\begin{figure}[p]
\epsfxsize=14.8cm
\epsfbox{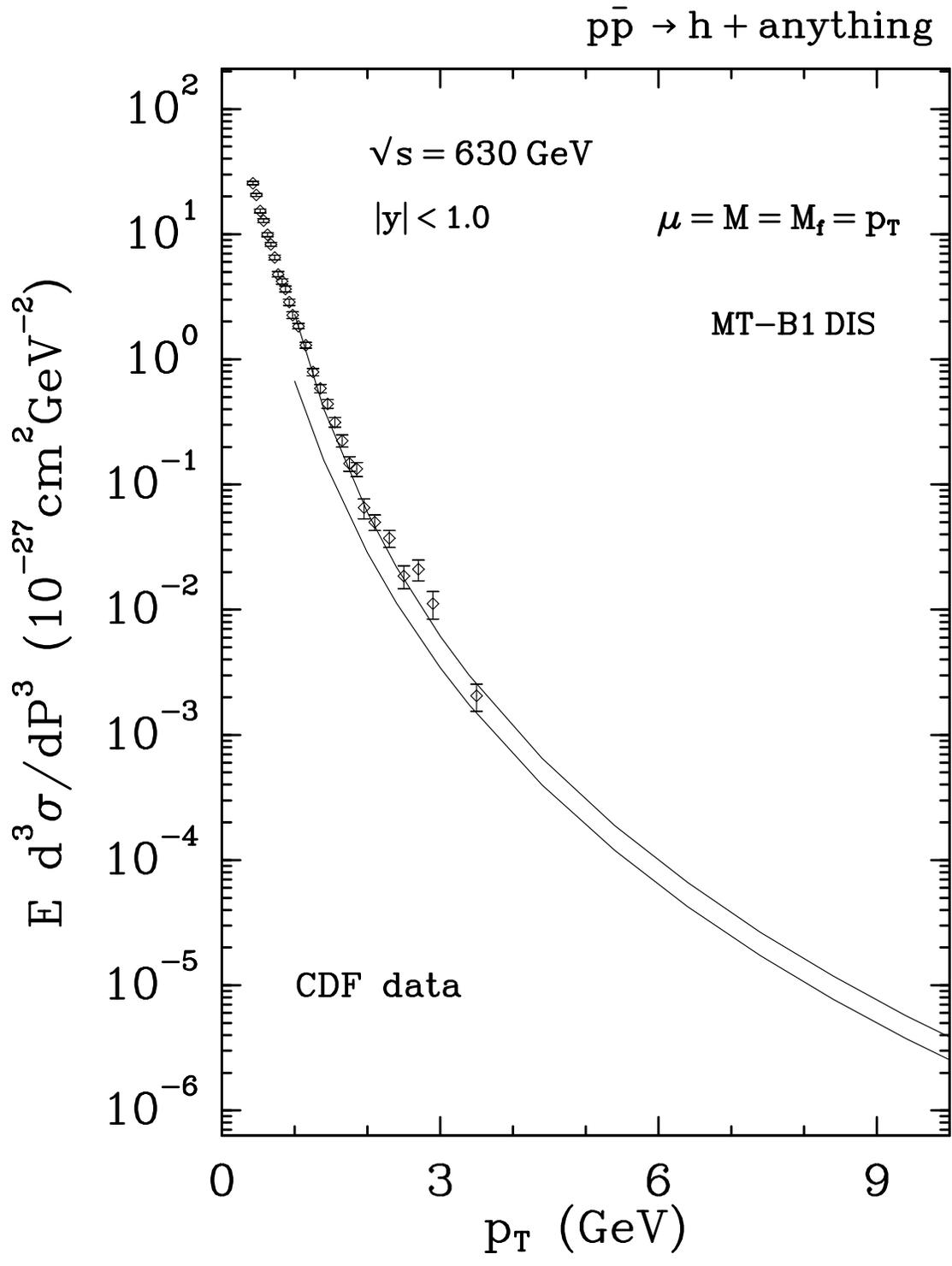}
\caption[f9]{\small{Same as in Fig.~\ref{UA2chb}
  for the data shown in Fig.~\ref{CDFonea}. }}
\label{CDFoneb}
\end{figure}
\begin{figure}[p]
\epsfxsize=14.8cm
\epsfbox{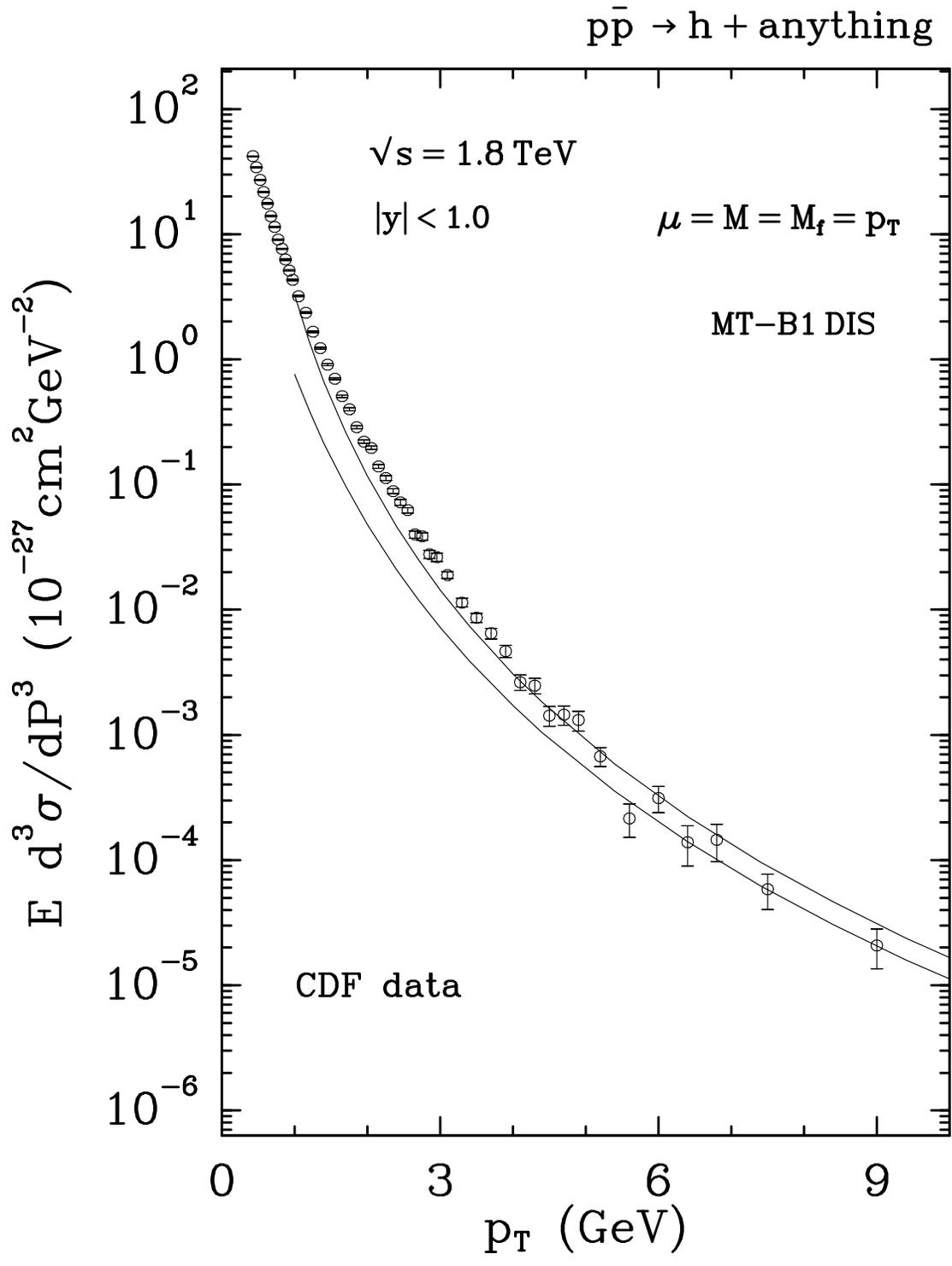}
\caption[f10]{\small{Same as in Fig.~\ref{UA2chb}
  for the data shown in Fig.~\ref{CDFtwoa}. }}
\label{CDFtwob}
\end{figure}
In order to get an idea of the size of the NLO correction in the hard
scattering terms, we compare in Figs.~\ref{UA2chb}, \ref{CDFoneb} and
\ref{CDFtwob} the Born cross section with the NLO one for
$\sqrt{s}=540,630\,$GeV and $1.8\,$TeV. Here, again, the cross sections
are averaged over the rapidity ranges in the same way as for the
respective experimental data. In the Born cross section the structure
functions as well as the value of $\as$ are the same as in the NLO
cross section. The effect of the NLO terms leads to an average K factor
of approximately $1.5$ for the choice of scales $\mu=M_f=M=\pT$.
For comparison, we show also the experimental data.

\begin{figure}[tb]
\epsfysize=12.4cm
\epsfbox{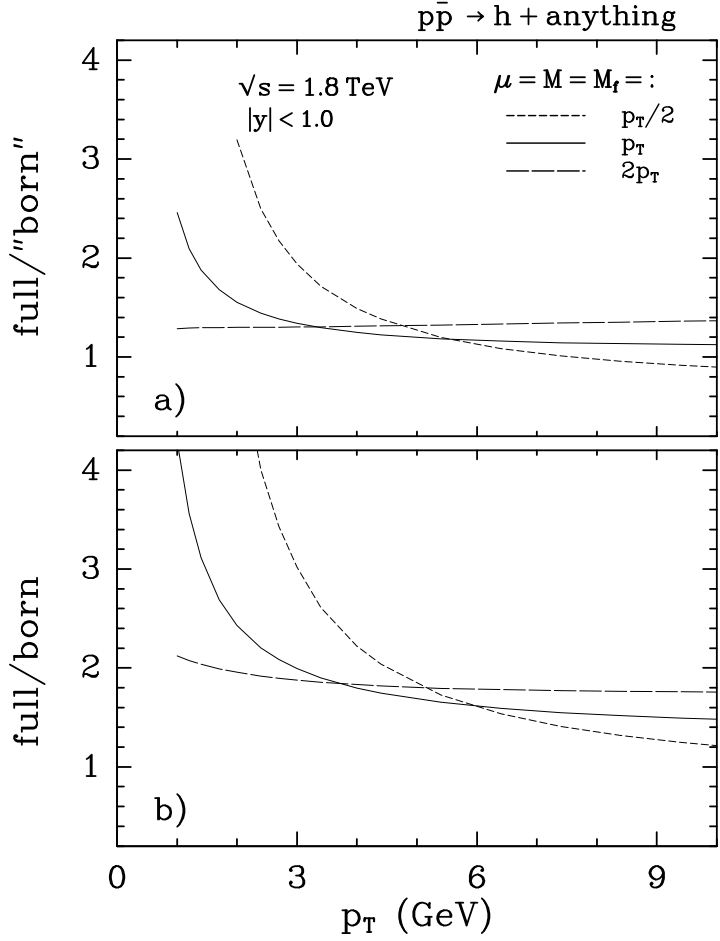}
\caption[f11]{\small{Ratios of the NLO charged-hadron
  production cross section over a) the LO one and b) the LO one
  evaluated with NLO structure functions and $\as$, for
  $\sqrt{s}=1.8\,$TeV, rapidity range $\vert y\vert<1.0$ and scales
  equal to $\pT/2$, $\pT$ and $2 \pT$.                  }}
\label{Kfactors}
\end{figure}
It has to be emphasized that this K factor is only a measure of the NLO
corrections to the hard scattering. Usually one takes the attitude of
comparing the NLO cross section to the LO one where the hard
scattering, the structure functions and the strong coupling constant
$\as$ are evaluated in lowest order. Such a K factor, denoted as
{\it full/``born''}, is presented in Fig.~\ref{Kfactors}a for the three
scales $\pT/2,\pT$ and $2 \pT$, CM energy $\sqrt{s}=1.8\,$TeV and
$\vert y\vert < 1.0$. The parametrization MT-SL is used for the
structure functions entering in the LO cross section. The value of
$144\,$MeV for the QCD parameter $\Lambda$ required by this
parametrization is extended also to the $\as$ coupling constant and to
the fragmentation functions. We observe in Fig.~\ref{Kfactors}a that
the K factor obtained in this way is $\sim 1$ and almost
$\pT$-independent for $\pT > 5\,$GeV if the three scales $\mu$, $M$ and
$M_f$ are set equal to $\pT$. For the other two scales, $\pT/2$ and $2
\pT$, the K factor has a somewhat stronger $\pT$ dependence and reaches
0.9 and 1.4, respectively, for the higher values of $\pT$, where
experimental data are available.

The K factor with structure functions and $\as$ kept at the NLO level,
referred to above ({\it full/born}), is shown in Fig.~\ref{Kfactors}b
also for the same choices of scales. It is larger than the one shown in
Fig.~\ref{Kfactors}a, ranging between 1.2 and 1.8 at the upper end of
the $\pT$-range studied here, and exhibits also a stronger scale and
$\pT$ dependence. We have verified that similar features are present
also at $\sqrt{s} = 540$ and $630\,$GeV. K factors between 1 and 2 are
obtained in these cases for scales varying between $\pT/2$ and
$2 \pT$, for $5\,$GeV$\!<\pT < 10\,$GeV. A weaker $\pT$ dependence is
observed at these lower CM energies: the ratios {\it full/born} are
in these cases essentially flat for $\pT>5\,$GeV as also appears in
Fig.~\ref{ratio}, where $\delta$-function fragmentation functions are
assumed.

It is clear from these figures that if one goes from a {\it bona fide}
LO theory to a NLO combination of hard-scattering cross sections,
structure functions and $\as$, a strong compensation between these NLO
effects takes place. On the other hand, less compensation is present
when the NLO effects are contained only in the hard scattering. The
importance of including the NLO corrections consistently in all
components, as already emphasized by many authors (see for example
\cite{JFO}), is therefore quite evident.

\begin{figure}[p]
\epsfxsize=13.6cm
\epsfbox{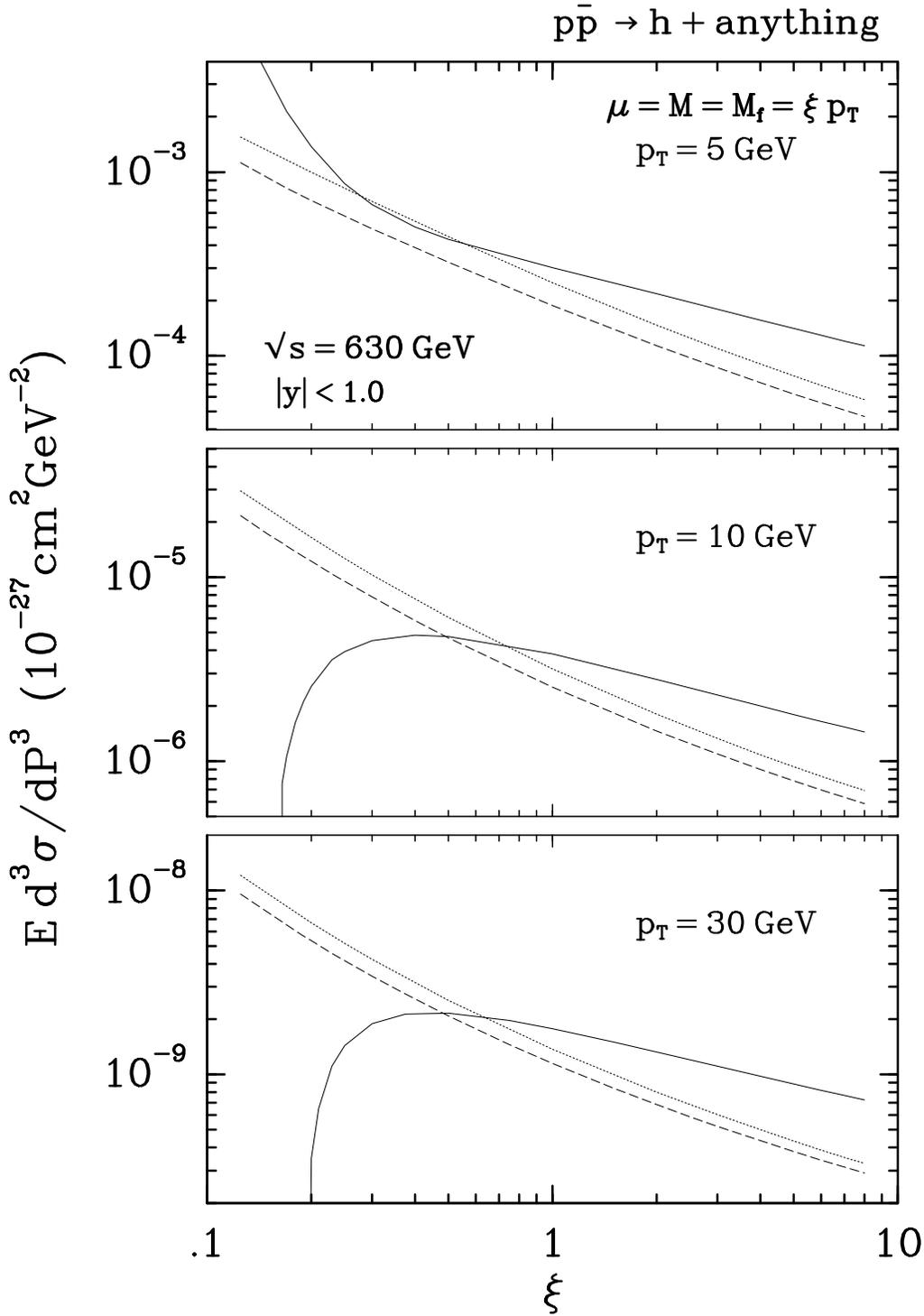}
\caption[f12]{\small{Scale dependence of the charged-hadron
  production cross section for $\sqrt{s}=630\,$GeV, $\vert y\vert<1.0$
  and $\pT =5, 10$ and $30\,$GeV. The solid lines show the scale
  dependence of the NLO cross section. The dotted and dashed lines
  show the behaviour of the {\it bona fide} LO cross section and
  the LO one with NLO structure functions and $\as$, respectively.
                           }}
\label{Scaledep}
\end{figure}
To obtain more information on the scale dependence of the inclusive
cross section, we have calculated it as a function of the scale factor
$\xi$ defined above for $\pT=5,10$ and $30\,$GeV, for CM energy
$\sqrt{s}=630\,$GeV and rapidity range $\vert y\vert < 1.0$. Results
are shown in Fig.~\ref{Scaledep} for the LO cross section (with LO
structure function and $\as$), for the NLO cross section and for a LO
cross section with the same structure functions and $\as$ used in the
NLO calculation.

We see quite clearly that the NLO cross section has much less scale
dependence than the two LO ones. The NLO cross section develops a
plateau at small values of $\xi$, for not-too-small $\pT$, i.e.
$\pT>5\,$GeV. The large turn-over of this plateau, however, is
presumably an artifact due to the use of LO fragmentation functions.
For all $\pT$, both LO curves show a monotonic dependence on $\xi$.
Between $\xi=1$ and $\xi=8$ the decrease of these two curves is steeper
than the one shown by the NLO cross section. We also observe that the
LO cross section (dotted curve) equals the NLO cross section for all
three $\pT$ at $\xi \approx 0.7$, whereas at larger values of $\xi$ the
two curves are quite well separated. This scale, where the true LO and
the NLO results match, is advocated as the best possible choice of
scales by the method of Fastest Apparent Convergence \cite{GRUNB}.

At this point, we should compare Fig.~\ref{Scaledep} for
$\pT=30\,$GeV with Fig.~\ref{scalpart}a, where the scale dependence has
been investigated at $\sqrt s=630$~GeV and $p_T=31.5$~GeV in the case
of $\delta$-function fragmentation. We notice that, apart from the
absolute size, the LO results exhibit roughly the same qualitative
features, which can be attributed to a rather weak scale dependence of
the fragmentation functions. However, the situation changes appreciably
when NLO corrections are included. Then, in the case of realistic
fragmentation the cross section falls off less steeply at large $\xi$,
due to compensations between the $M_f$ dependence of the fragmentation
functions and the $M_f$ dependence of the hard-scattering cross
sections. Finally, the drop-off at small $\xi$, when present, is also
more pronounced in Fig.~\ref{Scaledep}.

\begin{figure}[p]
\epsfxsize=13.6cm
\epsfbox{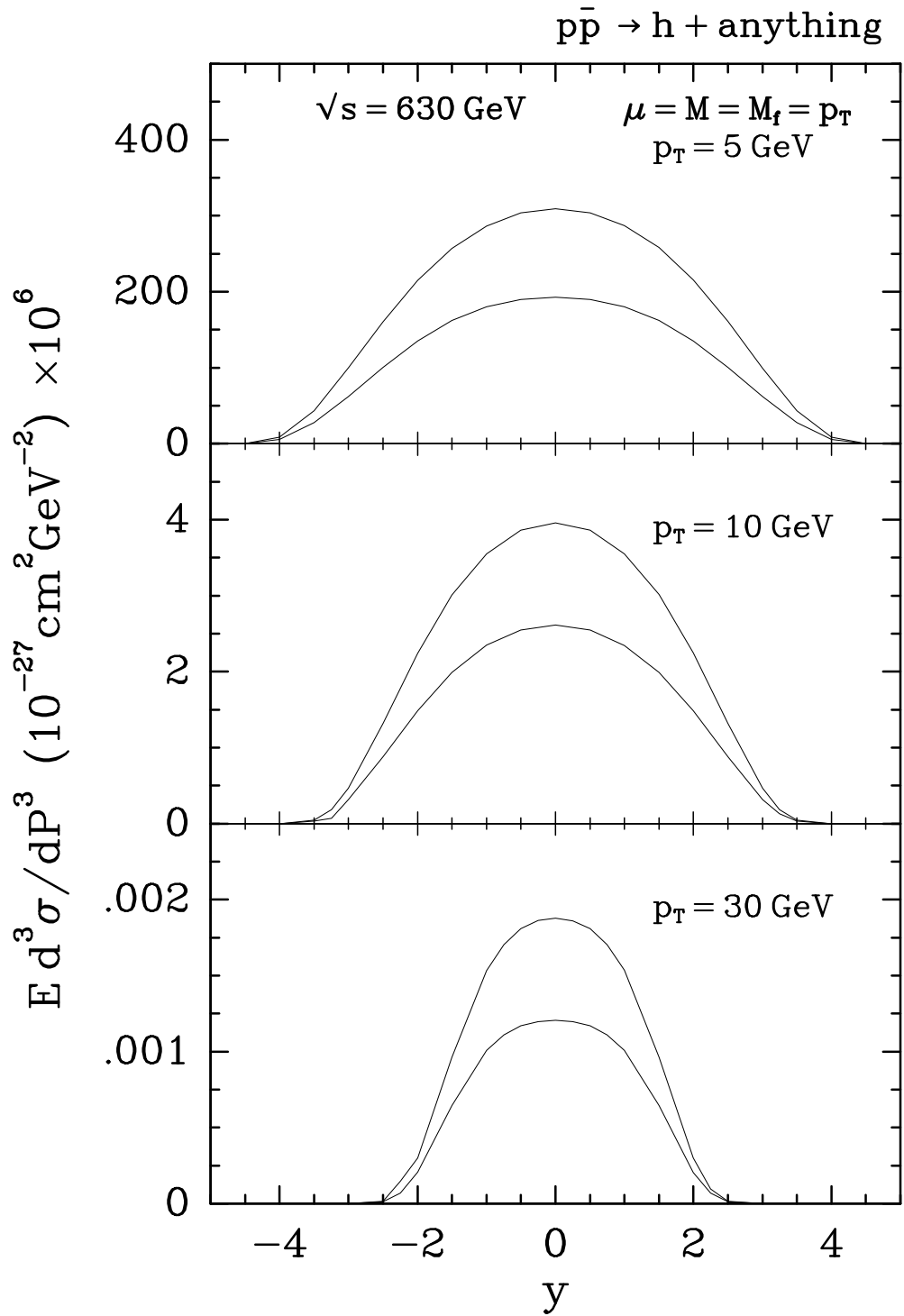}
\caption[f13]{\small{Rapidity distribution of the
  charged-hadron production cross section at $\sqrt{s}=630\,$GeV,
  scales fixed to $\pT$, for $\pT =5, 10$ and $30\,$GeV. The upper
  curves show the $y$ shape of the NLO cross section, the lower curves
  the shape of the LO one with structure functions and $\as$ at the NLO
  level. }}
\label{Rapidity}
\end{figure}
So far we have concentrated on the discussion of $\pT$ distributions
with rapidity integrated over a finite region. In
Fig.~\ref{Rapidity} we show the differential cross section
$d^3\sigma/d^2\pT dy$ as a function of the rapidity $y$ for CM energy
$\sqrt{s}=630\,$GeV and three different values of $\pT$:
$\pT=5,10$ and $30\,$GeV. We observe that the NLO corrections to the
hard-scattering cross section do not change the shape of this
distribution. The full NLO cross section and the LO one both exhibit
the same symmetric behaviour in $y$.

\begin{figure}[p]
\epsfxsize=14.8cm
\epsfbox{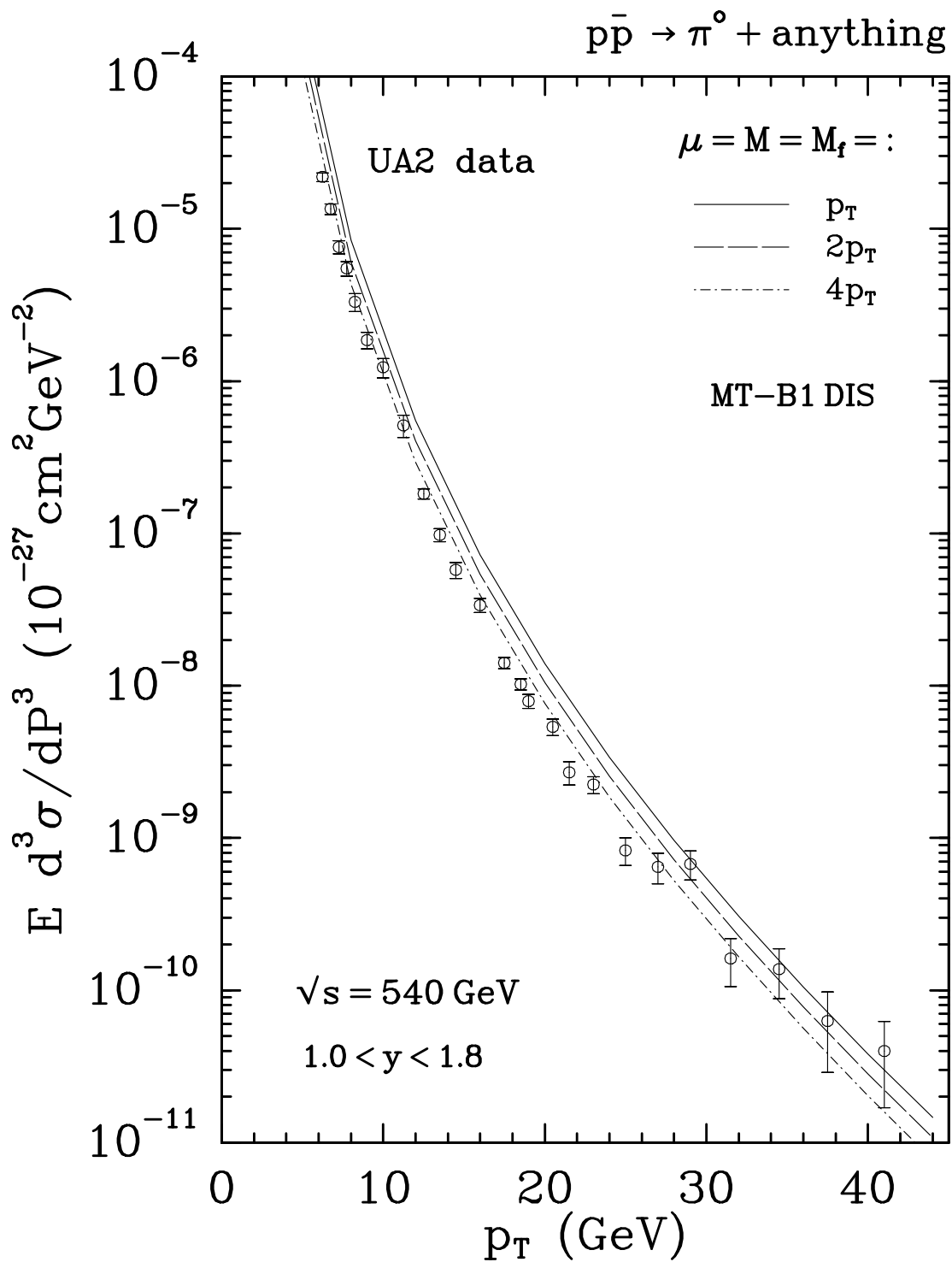}
\caption[f14]{\small{Inclusive $\pi^0$ production cross section for
  $\sqrt{s} =540\,$GeV averaged over the rapidity range
  $1.0 <y<1.8$ compared to the experimental data obtained by the UA2
  Collaboration. The short-dashed, solid and long-dashed lines
  correspond to the full NLO predictions for scales $\mu$, $M$ and
  $M_f$ all set equal to $\pT$, $2 \pT$ and $4\pT$, respectively.
                           }}
\label{UA2na}
\end{figure}
\begin{figure}[p]
\epsfxsize=14.8cm
\epsfbox{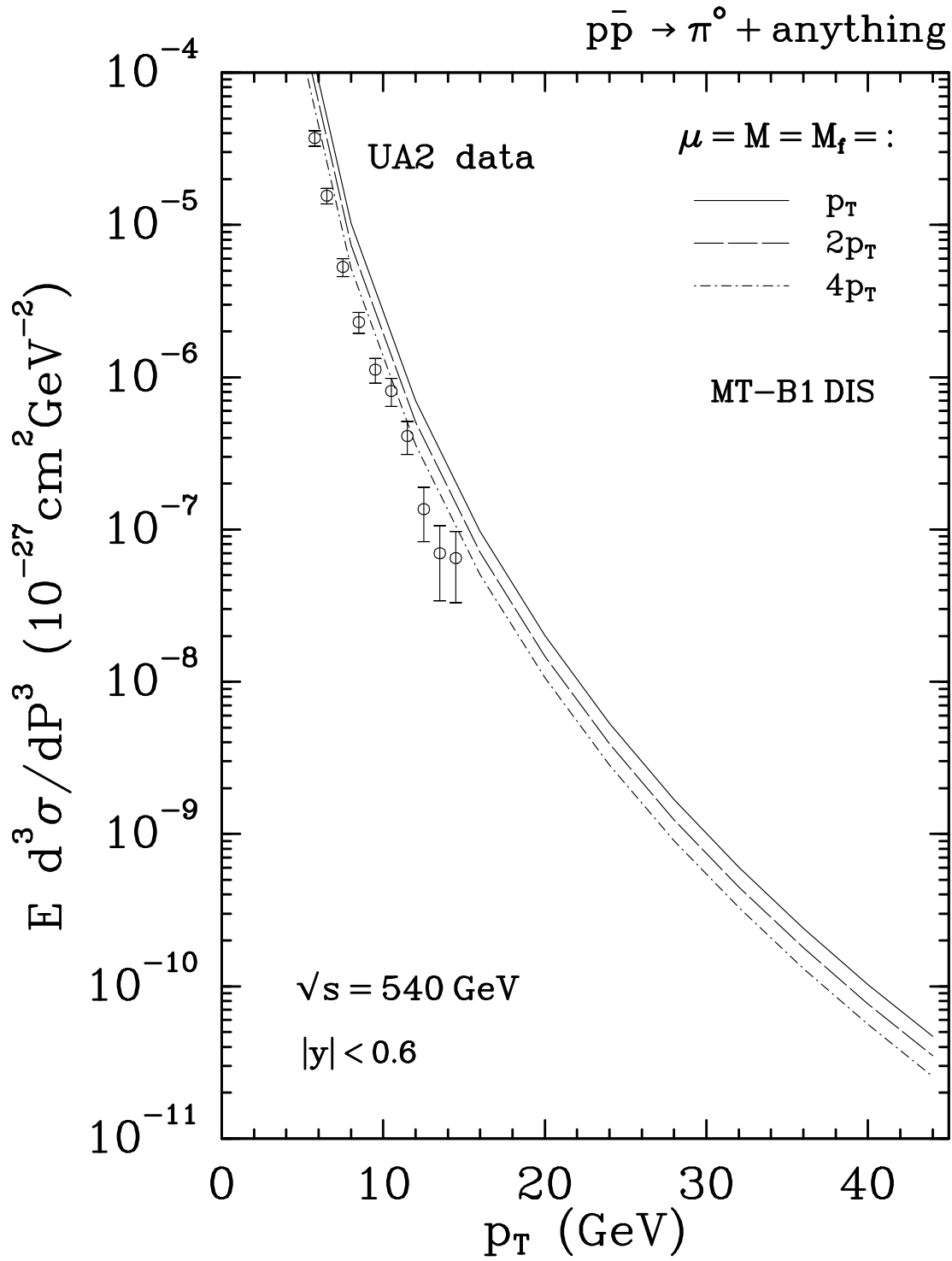}
\caption[f15]{\small{Same as in Fig.~\ref{UA2na} for rapidity
  range $\vert y\vert <1.0 $. }}
\label{UA2nb}
\end{figure}
At last, we show the results we obtain on inclusive $\pi^0$ production
at $\sqrt{s}=540\,$GeV for $1.0 <y<1.8$ (Fig.~\ref{UA2na}) and
$\vert y \vert <0.6$, compared with the UA2 data taken at the same CM
energy and in the same regions of rapidity (Fig.~\ref{UA2nb}).
It turns out that the data cannot be explained in NLO by using $\pT$
as scale, unlike the case of inclusive charged-hadron production. Both
choices of scales $\mu=M=M_f=\pT$ and $2\pT$ give cross sections
definitely above the data. A reasonably good fit can be obtained only
if we increase the value of these scales up to $4\pT$.
                                On the other hand, the fragmentation
functions into $\pi^0$'s might need to be revised, in particular the
gluon fragmentation function, which is not much constrained by the
$e^+e^-$ data used in \cite{BAI}. More work on the $\pi^0$
fragmentation functions is needed before a final conclusion concerning
the scale appropriate to the $\pi^0$ production cross section can be
drawn.
\newpage
\section{Summary and Conclusions}

We have calculated inclusive single-charged-hadron and
single-$\pi^0$ cross sections in NLO and compared our results with
experimental data from the UA2 and CDF Collaborations. The overall
agreement is satisfactory, surprisingly, also in the small-$\pT$
region. At large $\pT$, where QCD could be tested
cleanly, experimental data are unfortunately rare and plagued by
large errors. More accurate measurements in this region would
obviously provide a better probe of the QCD predictions.

As expected, the scale dependence in NLO is reduced
with respect to the one present in LO.
                Therefore, within the scale variation in NLO, our
results can be considered as an absolute prediction for
single-charged- and single-neutral-hadron production in $p \bar p$
collisions. The scale which yields the best agreement of the inclusive
charged-hadron production cross section to the experimental data
is equal to $\pT$. Larger values, up to $4 \pT$, are needed
to fit the data for inclusive $\pi^0$ production.

A drawback of our approach is that we still use fragmentation
functions in LO. While this will be improved in the future when
NLO fragmentation functions become available, we observe that
already now the scale dependence of our results is reduced with
respect to the one present in the LO calculation. We deduce
that some compensation concerning the factorization scale $M_f$
already works to a large extent.

When comparing our results to the LO ones with LO structure functions
and one-loop $\as$, we find that they roughly agree at scales just
below $\pT$. We consider this agreement as accidental, since they
deviate from each other at different scales. The fact that the NLO hard
scattering contributions are very important can be clearly seen when
the NLO results are compared with the LO ones with $\as$ and structure
functions taken in NLO. In this case, for scales equal to $\pT$, we
find a K factor approximately 1.5.

We plan to extend this NLO approach to inclusive hadron production in
$\gamma p$ and $e p$ collisions. We expect in this case a somewhat
different pattern due to the presence of the direct component in the
photon structure function. Further work is also needed in the
improvement of the fragmentation functions where a thorough analysis,
similar to the one performed for the structure functions, is
called for.

\vskip 3ex
\noindent
{\bf Acknowledgements}
We thank J.Ph.~Guillet for providing the Fortran implementation of the
results of \cite{AVE} together with clarifying comments and
S. Salesch for useful discussions. One of us (F.B.) thanks
D.\ Michelsen and H.\ Spiesberger for advice on the phase space Monte
Carlo integration and T.\ Ohl and M.\ Speh for support in using Emacs
under HP-UX.
\newpage

\end{document}